\documentstyle[prl,aps]{revtex}
\begin{document}
\draft
\flushbottom
\twocolumn[\hsize\textwidth\columnwidth\hsize\csname@twocolumnfalse\endcsname
\title{Implications of Charge Ordering for Single-Particle Properties of \\ 
High-$T_c$ Superconductors}

\author{ M.~I. Salkola$^1$, V.~J. Emery$^2$, and S.~A. Kivelson$^{3*}$}

\address{
$^1$Department of Physics and Astronomy, McMaster University, Hamilton, Ontario 
L8S 4M1, Canada \\
$^2$Department of Physics, Brookhaven National Laboratory,
Upton, New York 11973-5000 \\
$^3$Institute for Theoretical Physics, University of California,
Santa Barbara, California 93106
}
\date{February 5, 1996}

\maketitle
\tightenlines
\widetext
\advance\leftskip by 57pt
\advance\rightskip by 57pt
\begin{abstract}

The consequences of disordered charge stripes and antiphase spin
domains for the properties of the high-temperature superconductors
are studied. We focus on angle-resolved photoemission spectroscopy 
and optical conductivity, and show that the many unusual features
of the experimentally observed spectra can be understood naturally 
in this way. This interpretation of the data, when combined with 
evidence from neutron scattering and NMR, suggests that disordered
and fluctuating stripe phases are a common feature of high-temperature
superconductors.

\

\

\end{abstract}

]

\narrowtext
\tightenlines

Recent neutron-scattering experiments by Tranquada {\it et al.}
\cite{tranquada} have shown that the suppression of
superconductivity in La$_{1.6-x}$Nd$_{0.4}$Sr$_{0.12}$CuO$_4$
is associated with the formation
of an ordered array of charged stripes which are also 
antiphase domain walls between antiferromagnetically
ordered spins in the CuO$_2$ planes.
This observation strongly supports the
idea that disordered or fluctuating stripe phases are of
central importance for the physics
of high-temperature superconductors \cite{ekla}.  In this Letter, we
show that single-particle properties of a disordered
stripe phase can account for the exotic features in the
spectral density measured by angle-resolved photoemission
spectroscopy (ARPES) in Bi$_2$Sr$_2$CaCu$_2$O$_{8+x}$ (for which the best data is
available).  In particular, we compute the
spectral density in a background of disordered stripes
and show that it reproduces the experimentally observed shape of the
Fermi surface, the existence of nearly dispersionless states at 
the Fermi energy (``flat bands''), and the appearance of weak additional 
states (``shadow bands'') \cite{dessau,aebi,ma,ding}, features which 
have no natural explanation within conventional band theory. 
Finally, we comment briefly on the implications of partially ordered
spin domains for NMR experiments.

The principal signature of the antiphase spin domains in La$_{1.6-x}$Nd$_{0.4}$Sr$_x$CuO$_4$
is a set of resolution-limited peaks in the magnetic structure factor
at wave vectors $({1 \over 2} \pm \epsilon, {1 \over 2})$ and
$({1 \over 2}, {1 \over 2} \pm \epsilon)$ 
\cite{tranquada,units}.  The associated charge stripes are indicated by peaks in 
the nuclear structure factor at wave vectors $(\pm 2\epsilon,0)$ and
$(0,\pm 2\epsilon)$.
{\it A posteriori}, it is natural to interpret
the {\it inelastic} peaks in the magnetic structure factor
previously observed \cite{cheong} at similar locations
in reciprocal space  in superconducting samples
of La$_{2-x}$Sr$_x$CuO$_{4}$ as evidence of ``extended domains'' \cite{domain} of stripe 
{\it fluctuations}, in which the stripes are oriented along vertical or 
horizontal Cu-O bond directions respectively. Indeed, any experiment, such
as ARPES, that might be sensitive to the existence of an extended domain 
structure should be re-examined from this point of view; this is an important 
feature of our interpretation of the data.

Two mechanisms for producing stripe phases have been suggested by
theories of doped
Mott-Hubbard insulators:
a Fermi-surface instability \cite{schulz}
and frustrated phase separation \cite{emery}.
The former relies on Fermi-surface nesting which leads to 
a reduced density of states, or a gap, at the Fermi energy.
In the latter mechanism, a competition between phase separation
({\it i.e.}, the tendency of an antiferromagnetic insulator to expel doped
holes) and the long-range part of the Coulomb interaction
leads to charge-ordered phases, and especially stripe phases, which
may be either ordered, quantum melted, or disordered by quenched
randomness \cite{chayes}. The charge forms an array of {\it metallic} stripes, 
whose period is determined by the energetics of phase separation and is 
unrelated to any nesting vector of the Fermi surface. The charge structures,
in turn, drive the modulation of the antiferromagnetic order.
The experiments\cite{tranquada} on La$_{1.6-x}$Nd$_{0.4}$Sr$_x$CuO$_4$ clearly favor 
the latter point of view. The
ordering wave vectors do not nest the Fermi surface, and the
ordered system has partially-filled hole bands associated with the stripes.
Moreover the magnetic peaks first develop {\it below} the charge-ordering 
temperature \cite{tranquada,gl}.
Our interpretation of the ARPES experiments  on Bi$_2$Sr$_2$CaCu$_2$O$_{8+x}$ lends
further support to this conclusion: nesting would lead to a
diminished density of states at the Fermi surface,
whereas we find an {\it increased} density of states
corresponding to the flat bands seen in the experiments.

Our objective is to determine a phenomenological band structure
for electrons moving in an effective potential generated by charge
stripes and antiphase spin domains. We do not propose to solve
a particular many-body model by Hartree-Fock theory; indeed, we have
found that this approximation seems to favor insulating stripes,
even if the long-range part of the Coulomb interaction is included. Rather
we assume a phenomenological one-body Hamiltonian:
\begin{equation}
H= -t \sum_{\langle ll'\rangle\sigma}
( c^\dagger_{l\sigma} c_{l'\sigma} + H.c.)
 + \sum_{l\sigma} V_{\sigma}({\bf R}_l) n_{l\sigma},
\label{eq:heff}
\end{equation}
where the first term is the nearest-neighbor hopping on a square lattice and
the second one describes the interaction with the effective stripe potential.
Here, $c_{l\sigma}$ annihilates an electron of spin $\sigma
=\pm$ at site
${\bf R}_l$ and $n_{l\sigma} = c^\dagger_{l\sigma}c_{l\sigma}$.
The effective potential is given by:
\begin{equation}
 V_{\sigma}({\bf R})
 = \rho({\bf R}) +  \sigma S({\bf R}) e^{i{\bf Q}\cdot {\bf R}},
\end{equation}
where ${\bf Q}= ({\pi \over a},{\pi \over a})$ and $a$ is the lattice spacing.
Specifically, for vertical stripes, we use the
forms $\rho(x,y)= \rho_0 \sum_n {\rm sech}[(x-x_n)/\xi_c]$ and
$S(x,y)= S_0 \prod_n \tanh[(x-x_n)/\xi_s]$, where ${\bf R}=(x,y)$, $x_n$ 
are fixed centers
of the stripes, and the parameters $\rho_0$, $S_0$, $\xi_c$, and
$\xi_s$
determine the amplitude of the charge and spin modulation and
whether the stripes are narrow or broad.

According to the usual interpretation \cite{cardona}, 
the measured photo-current in a photoemission experiment
is the product of the electronic spectral density 
$A_-({\bf k},\epsilon)$ for the removal of one electron from the system 
and a slowly varying matrix element which reflects
the photon polarization selection rules.  
This spectral density can be written as
$A_-({\bf k},\epsilon) = f(\epsilon) A({\bf k},\epsilon)$, where
$f(\epsilon)=1/[e^{(\epsilon-\epsilon_F)/k_BT}+1]$ is the Fermi function, 
$\epsilon_F$ is the Fermi energy, and 
$A({\bf k},\epsilon) = -(1 / \pi) {\rm Im} G({\bf k},\epsilon + i0^+)$ 
is the spectral function of the one-electron Green's function 
$G({\bf k},t) = - i
\langle T c_{{\bf k}\sigma}(t) c^\dagger_{{\bf k}\sigma}(0) \rangle$.

First, consider vertical stripes condensed into a regular array:
$\rho(x+\ell)=\rho(x)$ and $s(x+2\ell) = s(x)$, where $\ell$ is
the separation between vertical stripes.   Results will be presented for 
bond-centered 
stripes, $x_n= n\ell + a/2$ with $\ell/a$ integer, but they are 
largely insensitive to this
assumption.
For even $\ell/a$, the unit cell size is $(2\ell/a)\times 2$
so the band-structure is computed
by diagonalizing a $(4\ell/a)\times(4\ell/a)$ matrix for each ${\bf k}$-vector.
For illustrative purposes, we have used the parameters 
$\rho_0=-t/2$, $S_0=2t$, $\xi_c=a$, and $\xi_s=2a$, for which the 
ground-state of the Hamiltonian
in Eq.~(\ref{eq:heff}) solves Hartree-Fock self-consistency
conditions at small doping for the
Hubbard model with $U/t= 4$ -- 5. However,  
to make contact with the structure observed  by Tranquada {\it et
al.}\cite{tranquada} at $1 \over 8$ doping, we choose
$\ell/a= 4$, which does not minimize the Hartree-Fock energy.
The results are not very sensitive to the choice of parameters, 
so long as the stripes are not too narrow. 

Figure 1 shows the spectral density 
$A_-$ (integrated over an energy window $\Delta \epsilon=t/30$ about 
$\epsilon_F$) as a function of ${\bf k}$.
Clearly the general shape of the calculated Fermi surface is quite different
from that of the noninteracting system (which circles the $\Gamma$ point) 
\cite{notation}. The fine features of the Fermi surface reflect 
the energy gaps at points spanned by the wave vectors $({1 \over 2} \pm 
\epsilon, {1 \over 2})$ of the spin order and $(\pm 2\epsilon, 0)$ of the 
charge order, where $\epsilon=a/2\ell$: they are generated by the multiple 
foldings of the energy band in the first Brillouin zone by the effective 
stripe potential $V_{\sigma}({\bf R})$. Figure 1 also shows shadow bands ---
weak copies of the Fermi surface created by the local doubling of the 
unit cell in the regions between the stripes. 

In order to compare with ARPES experiments on superconducting materials,
the stripes must be disordered \cite{order}. Since 
the slow collective stripe motion
is not strongly influenced by the single-particle dynamics,
we consider a quenched random distribution of stripes,
which we expect to give essentially the correct band structure \cite{ks}.
Specifically, with Bi$_2$Sr$_2$CaCu$_2$O$_{8+x}$ in mind, we 
chose $15$\% doping and a mean stripe separation  ${\ell}/a = 4$. 
The ensemble of stripe locations was constructed by taking 
$x_{n+1}-x_n= {\ell} + \delta$, where the random variable $\delta$ is 
uniformly distributed between $-3a$ and $3a$. The spectral density was 
averaged over five realizations, and we assumed a non-zero temperature, 
$k_BT=t/10$, which further diminished finite size effects.
We have found that the results do not depend markedly on the choice of
ensemble, or the parameters in the effective potential and that the large 
lattices used in the calculation (linear dimension 184 sites)
are essentially self-averaging. In other words, our
results are robust consequences of a disordered stripe
array, and are largely independent of other details. (We have not 
investigated the effects of orientational disorder.)

Figure 2 summarizes the results by showing the ${\bf k}$-dependence of the spectral 
density at the Fermi energy, and the quasi-electron dispersion along the line
$\Gamma$-$\bar{\rm M}_1$-X/Y for a single, extended domain, with disordered
vertical stripes (running in the
$\Gamma$-$\bar{\rm M}_2$ direction).
Disordering the stripes has removed the fine details from the Fermi surface
leaving only one sheet which closely resembles the Fermi surface of Refs.
\cite{dessau} and \cite{ma}. In particular, near $\bar{\rm M}_1$, 
there is a high density of states and a truly flat ``band'' at the 
Fermi energy, extending towards the $\Gamma$ and X/Y points.
The flatness along the $\Gamma$-$\bar{\rm M}_1$ line is a consequence
of both the smearing of the energy gap structure seen in the ordered system,
and the localization of the electronic wave functions
in the direction perpendicular to the stripes.
The spectral density of the shadow band
is reduced so much that it no longer shows up on a linear scale,
although it would reappear on a logarithmic scale.
In fact, plotted on such a scale,
$A_-$ looks qualitatively like that of Ref.~\cite{aebi}. 
The effect of vertical stripes at $\bar{\rm M}_2$ is  completely different: 
band narrowing in a direction parallel to the stripes leads to an open Fermi 
surface.

A stripe phase, even a disordered one, breaks the four-fold 
rotational symmetry of the ideal CuO$_2$ plane and reflection symmetry 
through a plane at 45$^\circ$ to the Cu-O bond. However 
reflection symmetry through planes parallel and perpendicular
to the stripes, and the associated
selection rules on the polarization dependence
of the matrix elements are still obeyed.
Extended domains with horizontal stripes give rise to the same
structures, but rotated through 90$^{\circ}$. With
the electric field polarized along the $\bar{\rm M}_1$ direction,
the photoelectron intensity 
vanishes by symmetry in the $\bar{\rm M}_2$ direction, and hence the
observed spectrum will 
de-emphasize the horizontal stripes, for which there are no Fermi-surface
crossings near $\bar{\rm M}_1$.
The photoemission experiments of Dessau {\it et al.} 
\cite{dessau} were performed in this geometry, so it is reasonable
to compare them directly with the results for an extended
domain of vertical stripes shown in Fig.~2(a);
indeed, the theoretical and experimental results
look remarkably similar.

For a Fermi liquid, the signature of well-defined quasiparticles
is a spectral density  $A_-({\bf k},\epsilon)$ which
approximates a $\delta$-function of energy as the energy $\epsilon$
approaches $\epsilon_F$. In the present calculation, it is clear from
the energy-dependence of the spectral density that
there are no well-defined quasiparticle features
near the $\bar{\rm M}_1$ point. One consequence is that 
the optical conductivity, shown in Fig.~3,  has a rather
small weight in the Drude component, with most of the oscillator
strength appearing in a broad, "midgap'' peak
centered in the neighborhood of $\hbar \omega \sim t$, which
merges into a weak, high-energy continuum. 
Although the stripe-induced electronic structure can appear to be an 
asymmetric dispersionless quasiparticle band, as shown 
in Fig.~2(b), the lack of well-defined quasiparticles is consistent 
with a widely held view of the normal-state properties of the high-temperature 
superconductors \cite{quasip}. This has profound implications for d.c.~transport and
other low-energy two-particle properties.

In summary, Bragg scattering from ordered stripe phases has 
been observed in neutron scattering in non-superconducting
La$_{1.6-x}$Nd$_{0.4}$Sr$_x$CuO$_4$, and strong evidence for disordered
and/or fluctuating stripes can be derived from the similar
structures seen in the dynamic spin structure factor of La$_{2-x}$Sr$_x$CuO$_{4}$.
The assumption that there exist disordered, or slowly
fluctuating stripes in Bi$_2$Sr$_2$CaCu$_2$O$_{8+x}$ provides a natural
explanation for the unusual features of the ARPES
data, including the shape of the
Fermi surface and the regions
of flat bands. Since there are also theoretical reasons for
believing that such combined charge
and spin structures are the natural consequences of
frustrated phase separation in a doped antiferromagnet, it is
reasonable to look anew at a wide variety of experiments in
the high-temperature superconductors to see whether they can be better understood in
terms of the properties of extended domains with short-ranged
stripe order.  For instance,
since  stripes break the four-fold rotational
symmetry of the crystal, dramatic consequences can be expected
for any {\it local} experiment which is designed to determine the
symmetry of the order parameter of the superconducting state.
This perspective will be examined in the still broader 
context of a global theory of synthetic ``bad metals'' in a forthcoming
publication.  However, it is worth mentioning some of the
more straightforward implications here, for concreteness focusing
on La$_{2-x}$Sr$_x$CuO$_{4}$.

Since the stripes are charged, they are easily pinned by disorder.
Thus, if the temperature is not too high, we can think of the
system as a quenched disordered
array of stripes, which divides
the Cu-O plane into long thin regions, with weak antiphase
coupling between the intervening hole-deficient regions.  This picture
rationalizes the observation \cite{statt} that
NMR sees two distinct species of Cu nuclei, which we
would associate with those in a pinned
stripe and those between the stripes.
Since the antiphase coupling between regions is potentially frustrating, this
picture gives a microscopic justification for the observation
of a ``cluster-spin-glass'' phase in samples with $x< 15$~\% \cite{spinglass}.
There is, moreover, evidence that the creation of dilute metallic
stripes can account for the rapid suppression of the N{\`e}el temperature
for $x<2$~\% \cite{borsa}.  

We are grateful to A. Loeser, Z.-X. Shen, J. Tranquada, and B. Wells for stimulating
discussions.
This work was supported in part by the Division of Materials Science,
U. S. Department of Energy under Contracts No. DE-AC02-76CH00016 (Brookhaven)
and by NSF Grant No. DMR-93-12606 (UCLA). M.I.S.~was supported in part by 
the U.S. Department of Energy, Natural Sciences and Engineering
Research Council of Canada, and the Ontario Center for Materials Research.


\begin{references}
\bibitem{units}   Here, wave vectors are expressed in units of $2\pi$          
divided by the lattice spacing.
\bibitem{domain} Conventionally, the word ``domain'' is used in two distinct
senses. One refers to the region of spins between the charge stripes, the other
to larger regions in which one of a discrete set of degenerate ground states is
realized. In this paper, we use the name ``extended domain'' for the second
case. Such a domain might be macroscopic or mesoscopic.
\bibitem{gl} A simple argument using Ginsburg-Landau theory shows that,
if the order is driven by the charge, there should be two transitions,
with spin ordering occurring at the lower temperature. Recent neutron
scattering experiments support this picture. (J. Tranquada, private
communication.)
\bibitem{order} In principle, a superconducting material could have an 
ordered array of stripes. However, there is no long-range magnetic order in 
La$_{2-x}$Sr$_x$CuO$_{4}$ and no evidence of charge order in 
La$_{2-x}$Sr$_x$CuO$_{4}$ or Bi$_2$Sr$_2$CaCu$_2$O$_{8+x}$. Consequently
it is appropriate to assume disordered stripes in calculating the band 
structure. The agreement with experiment indicates that this is a
realistic assumption.
\end{references}
\end{document}